\documentclass[12pt,preprint]{aastex}

\shorttitle{$H_0$ from Lensing of SNe by Clusters}
\shortauthors{Bolton \& Burles}

\begin{document}

\title{Prospects for the Determination of $H_0$
       through Observation of Multiply Imaged
       Supernovae in Galaxy Cluster Fields}

\author{Adam S. Bolton and Scott Burles}
\affil{Department of Physics and Center for Space Research \\
    Massachusetts Institute of Technology \\
    77 Massachusetts Avenue, Cambridge, MA 02139}
\email{bolton@mit.edu, burles@mit.edu}

\begin{abstract}
We assess the possibility of determining
the Hubble constant $H_0$ by measuring
time delays between multiple images
of supernovae gravitationally lensed
by rich clusters of galaxies and combining these
delay measurements with \mbox{detailed} cluster-potential
models based on other lensing constraints.
Such a lensing \mbox{determination} of $H_0$ would be
complementary to those obtained
from galaxy-QSO lensing studies,
and could potentially be better calibrated.
We show that relatively low-redshift
($z \sim 0.2$), significantly elliptical clusters
have appreciable lensing
cross sections for observable image pairings
with tractable time delays on the order
of a few years despite large lensing mass scales.
We find that a targeted search for
such image pairs would be a significant
undertaking for current observatories, but that it
would be appropriate for a facility such as the
proposed Large-aperture Synoptic Survey Telescope.
\end{abstract}

\keywords{distance scale---gravitational lensing---galaxies:
    clusters: general---supernovae: general}

\section{Introduction}

The best measurements of the Hubble constant
($H_0$) based on the Cepheid-calibrated distance
ladder and on the cosmic microwave background
agree impressively with one another:
$H_0 = 72 \pm 8 \, \mathrm{km} \, \mathrm{s}^{-1} \,
\mathrm{Mpc}^{-1}$ and $H_0 = 72 \pm 5 \, \mathrm{km}
\, \mathrm{s}^{-1} \, \mathrm{Mpc}^{-1}$,
respectively \citep{fre01, spe03}.  However,
these values are inconsistent with
much lower values
of $H_0$ based upon measurements
of gravitational lens
time delays and the assumption
that lensing galaxies
have extended dark halos \citep{koc02a, koc02b, koc03}.
In this paper we
explore the possibility of determining $H_0$ by
monitoring known strong-lensing galaxy clusters
and measuring time delays between
multiple images of high-redshift supernovae (SNe).

The number of known giant-arc clusters is
encouraging.
\citet{wu98} present a summary of 38 strongly lensing
clusters from the literature,
containing a total of 48 giant arcs and
arclets (see also \citet{wnb99}).  More recently,
\citet{lup99} have reported the discovery of
8 more giant-arc clusters, \citet{gye02} 6 more,
and \citet{zg03} 3 more.
Recent observations of the galaxy cluster
Abell 1689 with the Advanced Camera for Surveys (ACS)
on the {\sl Hubble Space Telescope} have revealed
at least
30 new multiply-imaged background sources, with
an average of $\sim$3 images each
(D. Coe, 9 January 2003 presentation
``Deep ACS and Keck Observations of A1689'', in
conference ``Gravitational Lensing: a Unique
Tool for Cosmology'', Aussois, Savoie, France).
All these image systems (together with previously
known strongly lensed features) provide constraints
on the projected gravitational potential of the
cluster---many more constraints than a quadruple-image
quasar lens system places on its lensing galaxy.
In addition, since multiple strongly lensed
sources behind a single galaxy cluster in
general lie at different redshifts, the cluster
potential map will not suffer the ``mass sheet degeneracy''
that afflicts galaxy-quasar lens systems \citep{sah00}.
Although A1689 is perhaps the strongest cluster lens in
the sky, we may assume that other known
strong-lensing clusters will likewise exhibit
many new multiple-image systems when observed
with ACS and later instruments, and detailed
cluster modeling like that of \citet{asw98a, asw98b}
and \citet{k95, k96} but with increased
accuracy will become possible.  Thus although
the structure of galaxy clusters is in general more
complex than that of individual galaxies, the projected 
potentials of clusters
may in the near future be more
tightly constrained by lensing than those of
individual galaxies, and
cluster-lens time delay would therefore
be more precisely calibrated.
(See \citet{sch00} for a general discussion of
the determination of $H_0$ from
gravitational-lens time delays.)

This work is distinguished from
previous studies by its
quantitative focus on the possibility of
measuring SN-image time delays in
cluster-scale lens systems,
and combining them with additional
strong-lensing constraints to determine $H_0$.
The original proposition to measure $H_0$
using gravitational lens time delays envisioned 
distant supernovae (SNe) as the variable
sources to be lensed by intervening galaxies \citep{ref64};
quasars and their variability had yet to be discovered.
\citet{kp88} examined the possibility of observing
SNe exploding within giant-arc source galaxies.
They made a detailed analysis for a
SN very near to and inside a caustic ``cusp''
and showed that for such configurations the
time delay between multiple images could be
as short as days to weeks, although they did not
explicitly consider the possibility of making an
$H_0$ determination.  \citet{gir92} reported
the detection of variability in two separate
arclets of the cluster Cl 0302+1658, consistent
with a single source and a time delay of
1 to 2 years.
Various papers have examined magnification
and time delay effects for SNe in field
surveys lensed by intervening halos across a
spectrum of mass scales
\citep{lsw88, mf98, df99, mfp00, pm00,
holz01, goo02, ost03, ok03},
and several have explicitly considered the
possibility of constraining $H_0$ using
SN time delays.  However, all these works
have modeled the lenses as circular,
and in this paper we show that circular
models seriously under-predict the
capacity for cluster-scale lenses to produce
multiple images with reasonably short time delays
(as one might anticipate based on the
results of \citet{kp88} for sources very
near to the caustic cusp of a lens).
Other studies have considered targeted
searches for lensed SNe in galaxy cluster
fields \citep{mr97, kb98, srs00, sul00, gms02, gg02},
but their quantitative focus has
been on pure SN detection rates and
not on the prospects for measuring time delays
that we quantify here.
With the exception of Saini et al.\ (2000), these
works have also approximated clusters with circularly
symmetric lens models.

This paper is organized as follows.
Section~\ref{theory} presents a brief
discussion of relevant gravitational lens
theory and defines the particular
lensing cross section that is central to
our calculations.
In Section~\ref{cross_section}, we examine
the lensing behavior of several cluster models
in the context of observational constraints on
minimum magnification and maximum time delay.
Section~\ref{accuracy} discusses the accuracy
in the determination of $H_0$ that we might
hope to obtain from individual
cluster time delays, both for a simplified
model and for realistic models.
Section~\ref{detect_sec} presents a calculation of
detection rates for observable
SN image pairings with acceptable time delays
in the strong-lensing region of a specific cluster.
Our approach in this section is similar to that
of \citet{sul00} and \citet{gg02}
(and differs from that of Saini et al. (2000)) in that
we do not directly consider the
possibility of SNe within known lensed galaxies,
but rather attempt a prediction
of SN detection rates in a targeted cluster
field based on based on an assumed 
cluster mass model and 
cosmic supernova rate density function.
(\citet{lan02} argue that above
redshifts of $\sim$1.5--2, even the Hubble Deep Field
observations are insensitive to the surface brightness
of most rest-frame ultraviolet emission.)
Finally, in Section~\ref{outlook}
we discuss the observational outlook in view of
the calculations of Section~\ref{detect_sec}.
We find that a measurement of $H_0$
based on cluster time delays would be
challenging but not impossible with today's
telescopes and instruments, and
that it would be an appropriate project for a large
telescope operating in a dedicated survey mode
such as the proposed Large-aperture Synoptic Survey
Telescope ({\sl LSST}) \citep{ty02}.

\enlargethispage*{1000pt}
Throughout this paper, we assume a flat,
vacuum-dominated cold dark matter ($\Lambda$CDM)
cosmology with matter and vacuum density parameters
$\Omega_{\mathrm{M}} = 0.3$, $\Omega_{\Lambda} = 0.7$.
For the Hubble constant, we take
$H_0 = 70 \, h_{70} \, \mathrm{km}
\, \mathrm{s}^{-1} \, \mathrm{Mpc}^{-1}$
with $h_{70} = 1$.
\pagebreak

\section{Lensing Theory}
\label{theory}

Gravitational lensing can be described elegantly through
the application of Fermat's
principle \citep{sch85, bn86}.
For a source at (2-vector) angular
position $\vec{\beta}$, the time of arrival
relative to the unlensed case for
image positions $\vec{\theta}$ is given by
\begin{equation}
\label{arrival_time}
t = {{(1 + z_{\mathrm{L}})} \over c}
{{D_{\mathrm{L}} D_{\mathrm{S}}} \over
{D_{\mathrm{LS}}}} \left[ \frac{1}{2}
(\vec{\theta} - \vec{\beta})^2
- \psi (\vec{\theta}) \right]~~,
\end{equation}
where $z_{\mathrm{L}}$ is the lens redshift, $D_{\mathrm{L}}$,
$D_{\mathrm{S}}$, and $D_{\mathrm{LS}}$ are angular diameter
distances to the lens, to the source, and from the lens
to the source, and $\psi (\vec{\theta})$
is proportional to the Newtonian potential of the lens
projected perpendicular to the line of sight \citep{nb96}.
Images of the source will be seen at positions $\vec{\theta}$
where the arrival time is stationary: that is, at
solutions to the ``lens equation'' obtained by setting the
gradient with respect to $\vec{\theta}$ of
(\ref{arrival_time}) to zero:
\begin{equation}
\label{lens_eq}
\vec{\beta}(\vec{\theta}) = \vec{\theta} -
\vec{\nabla}_{\vec{\theta}} \psi (\vec{\theta})~~.
\end{equation}
The scalar magnification $\mu$
(the flux ratio of lensed to unlensed
images of a source) is given by the ratio of lensed to
unlensed differential angular area, thus it is
given by the Jacobian determinant of the
mapping $\vec{\theta} (\vec{\beta})$
(the local inverse of (\ref{lens_eq})):
\begin{equation}
\mu = \det \left( {{d \theta_i} \over {d \beta_j}} \right)
= \left[ \det \left( {{d \beta_j} \over {d \theta_i}}
\right) \right]^{-1}~~.
\end{equation}
For an excellent presentation of gravitational
lensing in more detail,
we recommend the work of \citet{nb96}.

The detection of multiply imaged SNe will be limited
by both attainable photometric
depth and maximum tolerable time delay.
Accordingly, we define
$\sigma(\mu_{\mathrm{min}}, \Delta t_{\mathrm{max}})$
to be the angular cross section in the source plane
for lensing into a pair of images
with the fainter of the two having
(absolute-value) magnification of
at least $\mu_{\mathrm{min}}$ and the time delay
between the two being
at most $\Delta t_{\mathrm{max}}$.
For a particular physical lens model,
this cross section will depend upon
the redshifts of the lens and the source,
but we will suppress this dependence in our notation.

\section{Lens Models}
\label{cross_section}

In this section we examine
$\sigma(\mu_{\mathrm{min}}, \Delta t_{\mathrm{max}})$
for three progressively more detailed lens models,
and discuss the issue of
normalization for the purpose of comparing
between them.
These models
have been analyzed in detail by \citet{kk93}
and by Kormann, Schneider, and Bartelmann (1994).
Particularly convenient formulas
are given by \citet{kk98}.  We refer the
reader to these papers for analytic
lensing potential expressions
and for detailed discussions of the
models' lensing behavior.
Wherever numerical solution of the lens
equation (\ref{lens_eq}) is needed,
we use our own implementation of the grid-based
numerical lens-equation solution algorithm
with 2D Newton-method
solution refinement described by \citet{sef92}.

\subsection{Singular Isothermal Sphere}

The simplest reasonable model for an extended
astrophysical mass distribution is the familiar
singular isothermal sphere (SIS)\@.
It is parametrized solely by its
velocity dispersion $\sigma_v$.
The scaled, projected mass distribution
(convergence) of the SIS is given by
\begin{equation}
\kappa = {{\Sigma} \over {\Sigma_{\mathrm{cr}}}}
= {1 \over 2} {{\theta_{\mathrm{E}}} \over
{|\vec{\theta}|}} = {1 \over 2}
{{\theta_{\mathrm{E}}} \over
{\sqrt{\theta_x^2 + \theta_y^2}}} ~~.
\end{equation}
Here, $\Sigma$ is the physical surface mass density of
the lens and $\Sigma_{\mathrm{cr}} = (c^2 / 4 \pi G)
(D_{\mathrm{S}} / D_{\mathrm{L}} D_{\mathrm{LS}})$
is the so-called critical surface mass density.
$\theta_{\mathrm{E}}$ is the ``Einstein radius''
that sets the sole angular scale of the model
for given lens and source redshifts:
\begin{equation}
\label{theta_e}
\theta_{\mathrm{E}} =
4 \pi {{\sigma_v^2} \over {c^2}}
{{D_{\mathrm{LS}}} \over {D_{\mathrm{S}}}} \,
(\mathrm{radians}) =
(28.8 \, \mathrm{arcsec}) \left( {{\sigma_v} \over
{1000 \, \mathrm{km} / \mathrm{s}}} \right)^2
{{D_{\mathrm{LS}}} \over {D_{\mathrm{S}}}} ~~.
\end{equation}
The lensing potential of the SIS
is simply $\psi(\vec{\theta}) = \theta_{\mathrm{E}}
| \vec{\theta} |$.

We can obtain
$\sigma(\mu_{\mathrm{min}}, \Delta t_{\mathrm{max}})$
for the SIS analytically.
If the angular distance $\beta$ from the source (i.e., SN)
position to
the lens center is less than $\theta_{\mathrm{E}}$,
two images will be observed along a line on the sky
through the source position and the lens center: one at
a distance $\theta_{\mathrm{E}} + \beta$ from the lens
center (in the direction of the source) with magnification
$1 + \theta_{\mathrm{E}} / \beta$ and one
at a distance $\theta_{\mathrm{E}} - \beta$
(in the direction opposite the source)
with magnification $1 - \theta_{\mathrm{E}} / \beta$
\citep[for example]{nb96}.
This second, fainter image corresponds
to a saddle point of the arrival time function
(\ref{arrival_time}); its negative magnification
signals a reversal of image parity.  It is this
magnification that determines the $\mu_{\mathrm{min}}$
dependence of
$\sigma(\mu_{\mathrm{min}}, \Delta t_{\mathrm{max}})$
as defined.  If we substitute the solutions
$\theta_{\mathrm{E}} \pm \beta$ into (\ref{arrival_time})
(taking into account vectorial considerations)
and form the difference, we find that the time delay between
the two images is
\begin{equation}
\label{SIS_diff}
\Delta t =
{{2 (1 + z_{\mathrm{L}})} \over c}
{{D_{\mathrm{L}} D_{\mathrm{S}}} \over
{D_{\mathrm{LS}}}} \theta_{\mathrm{E}} \beta
\equiv 2 \tau {{\beta} \over {\theta_{\mathrm{E}}}}~~,
\end{equation}
where we have defined a characteristic lensing
timescale by $\tau$ as implied.  For a cluster
velocity dispersion $\sigma_v = 1000 \, \mathrm{km} / \mathrm{s}$,
$\tau$ increases steeply at first with increasing source
redshift, then levels off around $\sim25 \, h_{70}^{-1}$ years for
$z_{\mathrm{L}} = 0.1$ or $\sim$70--80$\, h_{70}^{-1}$ years
for $z_{\mathrm{L}} = 0.4$.  (In Section~\ref{sie_sec}
we will relieve our sense of discouragement over
such long timescales.)

We see that for the SIS a given $\beta$ corresponds to
a unique $\Delta t$ as given by (\ref{SIS_diff}) and a unique
(absolute-value) magnification of
$\theta_{\mathrm{E}} / \beta - 1$ for the fainter
image, so the form of
$\sigma(\mu_{\mathrm{min}}, \Delta t_{\mathrm{max}})$
is particularly simple:
we have
two singly limited cross sections given by
\begin{mathletters}
\begin{eqnarray}
\label{single_lim_m}
\sigma(\mu_{\mathrm{min}}) &=&
\pi \theta_{\mathrm{E}}^2
(\mu_{\mathrm{min}} + 1)^{-2} \\
\label{single_lim_t}
\sigma(\Delta t_{\mathrm{max}}) &=&
\frac{1}{4} \pi \theta_{\mathrm{E}}^2
(\Delta t_{\mathrm{max}} / \tau)^2~~,
\end{eqnarray}
\end{mathletters}
and
\begin{equation}
\label{SIS_lim}
\sigma(\mu_{\mathrm{min}}, \Delta t_{\mathrm{max}})
= \min \left[ \sigma(\mu_{\mathrm{min}}),
\sigma(\Delta t_{\mathrm{max}}) \right]~~.
\end{equation}
That is, for a given
$(\mu_{\mathrm{min}}, \Delta t_{\mathrm{max}})$,
we are either magnification-limited or delay-limited.
(\mbox{Arguably} the maximum possible double-image cross
section $\pi \theta_{\mathrm{E}}^2$ should be
enforced in (\ref{single_lim_t}), but it is built
into (\ref{single_lim_m}) and thereby propagates
to (\ref{SIS_lim}).)

\subsection{Singular Isothermal Ellipsoid}
\label{sie_sec}

By making the isodensity contours of the SIS
elliptical with a minor-to-major axis
ratio $q$, we get the singular isothermal ellipsoid
(SIE) model:
\begin{equation}
\label{SIE}
\kappa = {1 \over 2}
{{\theta_{\mathrm{E}}} \over
{\sqrt{q \theta_x^2 + q^{-1} \theta_y^2}}} ~~.
\end{equation}
As noted by Kormann et al.\ (1994), the total
mass enclosed within a given isodensity contour
remains constant with changing $q$ at fixed
$\theta_{\mathrm{E}}$ when $\kappa$ is expressed
in this form.  We adopt this normalization as
a basis for comparing models of differing
axis ratios, with $\theta_{\mathrm{E}}$ still
given as a function of $\sigma_v$ and
source and lens redshifts by (\ref{theta_e}).

The introduction of ellipticity leads to richer
lensing phenomena.  Most significantly, we acquire
a cross section for quadruple imaging.
For axis ratios $q$ less than 1 but
greater than about 0.394,
four images of a source will form if
the source position lies within a
diamond-shaped ``tangential'' caustic
surrounding the lens center and inside
the original border between singly and doubly
imaged regions (the ``radial cut'').
Labeling these
quad images in order of increasing arrival time,
1 and 2 are minima
of the arrival time function and
3 and 4 are saddle points.  2 and 3 are
in general of much greater absolute magnification
than 1 and 4.  We now have the possibility
of three independent image pairings.
For $q < \approx 0.394$, the major-axis
cusps of the tangential caustic extend outside
the radial cut and are referred to as
``naked cusps''; sources within the cusps
but outside the radial cut will be triply imaged.
\citet{kk93} Figure 1(d,e)
illustrates naked and non-naked cusps.

\begin{figure}
\plotone{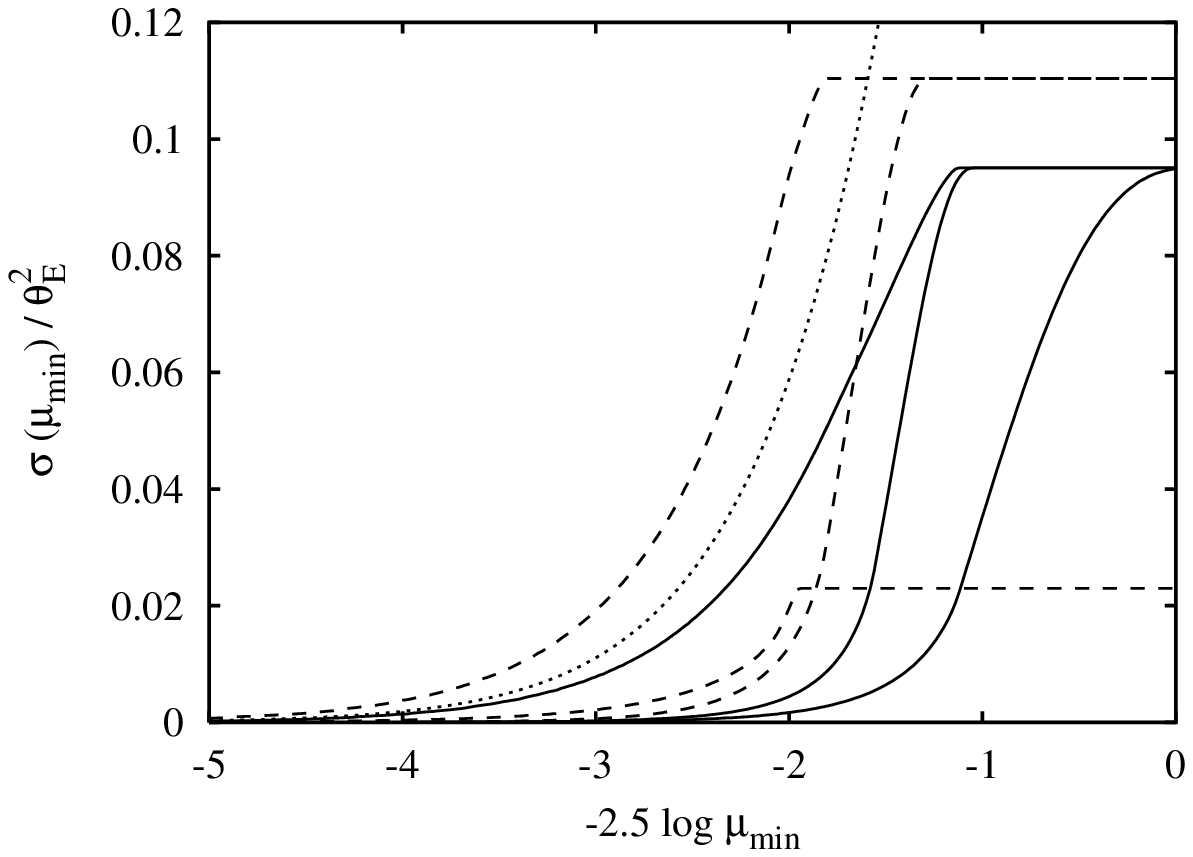}
\caption{Magnification-limited lensing cross
sections.  Solid lines are for a $q = 0.65$ SIE
quad.  The upper, middle, and lower curves correspond to the
2nd/3rd, 1st/2nd, and 3rd/4th image
pairings respectively.  Dashed lines are for the
corresponding images of a
$q = 0.65$ NIE at the ``umbilic catastrophe''
and with the normalization
$\theta_{\mathrm{E}} = \overline{\theta}_{\mathrm{E}}$
as described in the text.
(The curve with the reduced maximum cross
section is for the 3rd/4th image pairing.)
The dotted curve is for the two
images of a $q = 1$ SIS, for comparison.
\label{sigma_m}}
\end{figure}

\begin{figure}
\plotone{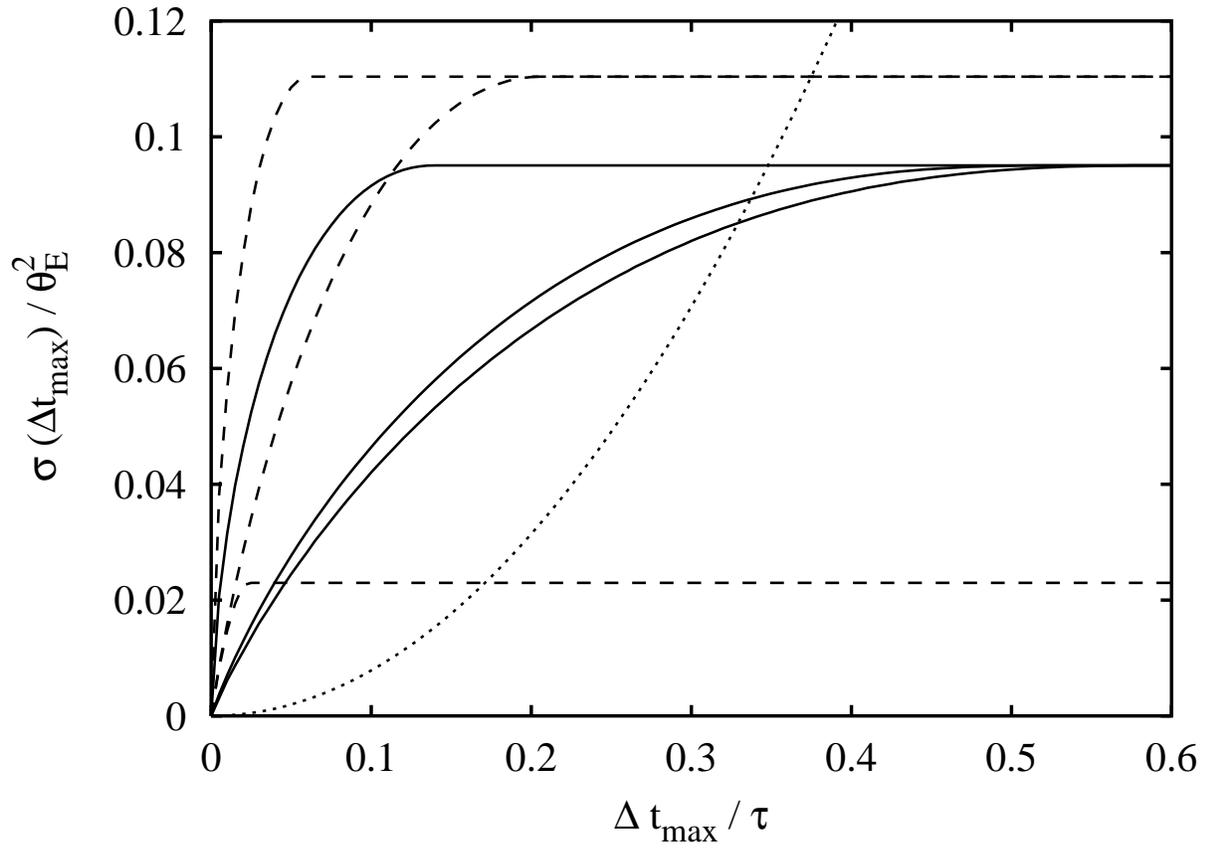}
\caption{Same as Figure~\ref{sigma_m}, but for
time-delay-limited lensing cross sections.
\label{sigma_t}}
\end{figure}

The SIE model has an angular scale invariance
that allows us to generate the function
$\sigma(\mu_{\mathrm{min}}, \Delta t_{\mathrm{max}})$
once for a given $q$, and rescale it as needed
for any source/lens redshift combination.
With circular symmetry broken,
we would not necessarily expect the
doubly limited cross sections for the various
image pairings to be of the form
(\ref{SIS_lim}), but our numerical calculations
show that such a form in fact yields a
very good approximation.
The solid curves in
Figures~\ref{sigma_m} and~\ref{sigma_t} show
$\sigma(\mu_{\mathrm{min}})$
and $\sigma(\Delta t_{\mathrm{max}})$
for the three independent quad image pairings
of least time delay for an SIE lens model with
axis ratio $q = 0.65$.
$\sigma(\mu_{\mathrm{min}}, \Delta t_{\mathrm{max}})$
can then be constructed for each image pairing
as per (\ref{SIS_lim}).  For a given $q$,
$\sigma(\mu_{\mathrm{min}}, \Delta t_{\mathrm{max}})$
of any quad image pairing is limited
by the full angular area of the
region in the source plane enclosed by the
tangential caustic.  This area increases as the
lens becomes more elliptical,
and is approximately $0.095\, \theta_{\mathrm{E}}^2$
for $q = 0.65$ as seen in Figures~\ref{sigma_m}
and~\ref{sigma_t}.

By comparing the solid SIE curves of Figures~\ref{sigma_m}
and~\ref{sigma_t} to the dotted SIS double-imaging curve,
we see that that the introduction
of ellipticity significantly increases the
cross section for multiple-image lensing with relatively
short time delays as compared to the
circularly symmetric SIS-lens case, at the cost of
reduced fainter-image magnification
\footnote{Here we are comparing the high-magnification
behavior of a circular lens to that of an
elliptical lens.  For any
{\em particular} elliptical lens, a quad
configuration will be of greater total magnification
than a double, since the quadruple-imaging cross
section carves out the very highly magnified
region of the source plane directly behind the lens.
Hence observed quad-image QSO
lenses generally have greater magnification than
doubles.}.
Of the three independent
image pairings in the quad case,
the 2nd/3rd image pairing is most significant for affording
both large magnification and short time delay;
these are the two images that
will merge and annihilate if the
source crosses outside the
tangential caustic.
When the source approaches the center of the
lens, the four images form a cross
and the 2nd/3rd image time delay approaches its maximum
value.  This limit can be found by solving the
lens equation in closed form for the special
case of a source directly behind the lens center:
\begin{equation}
\label{dtmax23}
\Delta t_{\mathrm{max, \,2 \, to \, 3}} =
{{\tau} \over {2 (q^{-1} - q)}}
\left( \mathrm{arctanh}^2 \sqrt{1 - q^2} -
\arctan^2 \sqrt{q^{-2} - 1} \right) ~~.
\end{equation}
For $q = 0.65$, this is about 0.14$\tau$ as seen
in Figures~\ref{sigma_m} and~\ref{sigma_t}.
For a lens redshift $z_{\mathrm{L}} < \sim 0.1$,
the lensing timescale $\tau$
as defined in (\ref{SIS_diff})
will be $< \sim 25 \, h_{70}^{-1}$ years for
a cluster velocity dispersion
$\sigma_v = 1000 \, \mathrm{km} / \mathrm{s}$,
and all 2nd/3rd image pairings
of a $q=0.65$ lens will have time delays
of less than $\sim 3.5 \, h_{70}^{-1}$ years.
Thus although sources very near
to the caustic will indeed have very short time delays
as shown by \citet{kp88}, sources {\em anywhere} within
the tangential caustic can still be lensed into multiple
images with observationally acceptable time delays.
Furthermore, as the 2nd/3rd image time delay approaches
its maximum, the 1st/2nd and 3rd/4th image time delays
approach zero for reasons of symmetry,
so (\ref{dtmax23}) is in fact only
an upper limit on the shortest
time delay experienced by any quadruply imaged source.

With all other parameters fixed, both $\tau$ and
$\theta_{\mathrm{E}}^2$ scale as the fourth power
of the cluster velocity dispersion $\sigma_v$.  Therefore
if the axes of Figure~\ref{sigma_t} were labeled
in units of years and square arcseconds (instead
of in units of $\tau$ and $\theta_{\mathrm{E}}^2$), an increase
in $\sigma_v$ would rescale both axes by the same factor.
This fact together with the differing concavities of the
curves shown in Figure~\ref{sigma_t} indicates that
in delay-limited cases, increasing $\sigma_v$ will
increase an SIE-quad cross section, but decrease an
SIS-double cross section.

\subsection{Nonsingular Isothermal Ellipsoid}
\label{NIE_sec}

A singular isothermal core is perhaps an unrealistic
feature to assume in a galaxy cluster.  To investigate
sub-isothermal core behavior, we can add a core
of angular radius $s$ to the SIE to
obtain the nonsingular isothermal ellipsoid (NIE):
\begin{equation}
\label{NIE}
\kappa = {1 \over 2}
{b \over {\sqrt{ s^2 +
q \theta_x^2 + q^{-1} \theta_y^2}}} ~~.
\end{equation}

To compare the lensing behavior of
models with differing core radii, we must take
care to adopt a sensible normalization.
In the limit $s \rightarrow 0$, $b$ may be identified
with $\theta_{\mathrm{E}}$ from above, but to compare
singular and nonsingular models the normalization
$b = \theta_{\mathrm{E}}$ is inappropriate.  Instead
we take
\begin{equation}
\label{norm}
b = {{\theta_{\mathrm{E}}} \over
{\overline{\theta}_{\mathrm{E}}}}
\left( sq + \sqrt{s^2 + \overline{\theta}_{\mathrm{E}}^2}
\right) ~~,
\end{equation}
where $\overline{\theta}_{\mathrm{E}}$
is the Einstein radius (\ref{theta_e})
evaluated for a source at some fiducial
redshift $\overline{z}_{\mathrm{S}}$.
With this normalization,
$\kappa$ retains the proper scaling with changing
source redshift, and the
tangential critical curve crosses the major
axis of the lens at fixed position
$\overline{\theta}_{\mathrm{E}} / \sqrt{q}$
with varying $s$ for a 
source at redshift $\overline{z}_{\mathrm{S}}$.
This choice is meant to respect
(approximately) the constraint that would be
placed on the cluster mass distribution by a giant arc at
redshift $\overline{z}_{\mathrm{S}}$.

Several changes occur when we increase the
core radius from zero.  First,
the model now has two angular scales, $b$ and $s$.
For a given lensing cluster, $b$ will change with varying
source redshift while $s$ remains fixed.  The
scale invariance of the lens is broken, and
the lensing behavior changes qualitatively between
source planes at different redshifts;
we now need to compute cross sections for
a range of $s / b$ and to scale them
appropriately.
Second, in multiple-imaging
configurations, an additional image that
was previously infinitely demagnified in the
singular core becomes in principle observable
(accordingly, the ``radial cut'' of the singular
case is now the ``radial caustic'').
This new image corresponds to a maximum of the arrival-time
function~(\ref{arrival_time}).  In this study we exclude
these ``maximum'' images from consideration so as to
have continuity of the total cross section
$\sigma(\mu_{\mathrm{min}}, \Delta t_{\mathrm{max}})$
from the singular case
when summing the contributions of all independent
image pairings.  (Actual core images could also
get lost in the light of a cluster cD galaxy.)
Finally, although the overall cross section for
multiple imaging decreases with increasing core radius
through a contraction of the radial caustic,
the area enclosed by the {\em tangential} caustic
actually {\em increases} slightly.
The decreased curvature of the arrival-time
surface in the central regions of the lens
also leads to larger magnifications and shorter time
delays.  Thus the cross section
$\sigma(\mu_{\mathrm{min}}, \Delta t_{\mathrm{max}})$
at fixed $\mu_{\mathrm{min}}$,
$\Delta t_{\mathrm{max}}$ tends to increase for all
image pairings as the core radius is increased,
until the formation of naked cusps begins
to reduce the cross section for the
3rd/4th image pairing.
The dashed lines in Figures~\ref{sigma_m} and~\ref{sigma_t}
show the cross sections for the same image pairings as
in the SIE case above, but for an NIE lens at
the so-called
``umbilic catastrophe'' and for sources at the
same redshift as the normalization redshift
$\overline{z}_{\mathrm{S}}$.
Here the umbilic catastrophe refers to
the particular $s$ value at which the expanding
tangential caustic completely engulfs the shrinking
radial caustic; see
\citet{kk93} Figure 1(c) for an illustration.
The NIE cross sections
for the 1st/2nd and 2nd/3rd image pairings are increased
over the SIE case at all values of $\mu_{\mathrm{min}}$,
$\Delta t_{\mathrm{max}}$.  All 2nd/3rd
image pairings now have time delays
$< \sim 0.06 \tau$.

\subsection{More Complicated Models}

The detailed lensing behavior of a
real galaxy cluster
cannot be captured entirely by the models
considered above.  The effects of
unmodeled cluster substructure are likely to
be somewhat similar to the effects
of increasing cluster ellipticity
relative to the circular case:
higher image multiplicity, shorter time delays between multiple
images, and reduced magnification.  In fact, ellipticity
may be thought of as the leading order of
substructure beyond circular symmetry.
By considering the perhaps exaggerated
axis ratio $q = 0.65$ we hope to
approximate unmodeled substructure effects.
An obvious refinement would be
to carry out our cross section calculations using a more
detailed and realistic cluster mass map,
but we defer this possibility to
future investigations.
Our primary goal is to
estimate the detection rates of multiply
imaged SNe in a single cluster field
that would be useful for the determination
of $H_0$, and thus we confine our attention to
the preceding models.

\section{Accuracy of $H_0$ Determination}
\label{accuracy}

First we will examine the usefulness of a
measured time delay in determining $H_0$
within the simplified context of the SIE lens model.
By the convenient result published by \citet{wmk00}
for self-similar isothermal lenses,
the time delay between
images A and B is given by
\begin{equation}
\label{SIEdelay}
\Delta t_{\mathrm{AB}} = {1 \over {2 c}}
{{D_{\mathrm{L}} D_{\mathrm{S}}} \over {D_{\mathrm{LS}}}}
(1 + z_{\mathrm{L}}) \left( |\vec{\theta}_{\mathrm{B}}|^2
- |\vec{\theta}_{\mathrm{A}}|^2 \right) ~~,
\end{equation}
with the image positions measured relative to the
lens center.  $c H_0^{-1}$ factors out of the
$D$'s on the right-hand side.
The dependence on the cosmological parameters
$(\Omega_{\mathrm{M}}, \Omega_{\Lambda})$
is generally weak but worth noting.
For example, taking
a lens redshift of 0.1 (0.3) and a source redshift of~1,
$D_{\mathrm{L}} D_{\mathrm{S}} / D_{\mathrm{LS}}$
changes only by about 1\% (3\%) as $\Omega_{\Lambda}$ goes
from 0 to 0.7 in flat universes.
(If we allow for unknown
cosmological density parameters, then our
time delay will not give us $H_0$,
but rather the distance measure
$D_{\mathrm{L}} D_{\mathrm{S}} / D_{\mathrm{LS}}$.)
Assuming the image positions, source
and lens redshifts, and
time delay can be measured with high
accuracy, the greatest source of error in
an $H_0$ determination is due to the
uncertainty $\delta \vec{\theta}_0$
in the position of the center of the lens.
From (\ref{SIEdelay}), this is
\begin{equation}
\label{delta_H}
{{\delta H_0} \over {H_0}} =
2 {{(\vec{\theta}_{\mathrm{B}} - \vec{\theta}_{\mathrm{A}})
\cdot \delta \vec{\theta}_0} \over
{|\vec{\theta}_{\mathrm{B}}|^2
- |\vec{\theta}_{\mathrm{A}}|^2}}
\equiv \vec{V} \cdot \delta \vec{\theta}_0
/ \theta_{\mathrm{E}} ~~.
\end{equation}
For positions of
2nd/3rd image pairings obtained numerically,
the dimensionless vector $\vec{V}$ ranges in magnitude
from 2 for source positions near the lens center
to almost zero for source positions just
inside the tangential caustic.
$\vec{V}$ is approximately of unit magnitude for
source positions of
area-weighted median fainter-image magnification.
The fractional error
in an $H_0$ determination is then
roughly equal to the uncertainty
(parallel to the image separation vector)
of the lens center position
in units of the Einstein angular
scale $\theta_{\mathrm{E}}$ defined in (\ref{theta_e})
and evaluated for the source redshift of the SN.
However, the magnification of both images increases
as the source approaches the tangential caustic,
so a magnification bias favors the detection of
events with smaller fractional
uncertainties in the derived $H_0$.

Of course, our real hope would be to
use not an SIE model (nor even an NIE model) but a
detailed lens model that takes
full account of lensing
constraints like giant arcs and multiply
imaged galaxies.
In this more general case,
the fractional uncertainty in
a derived $H_0$ value will be related to the
fractional uncertainty in the lensing potential
$\psi (\vec{\theta})$ and the square
of its gradient, evaluated
at the two SN image positions.
(This can be seen by
substituting (\ref{lens_eq}) into (\ref{arrival_time})
to eliminate the source position $\vec{\beta}$.)
Based on their nonparametric lensing inversions,
AbdelSalam et al.\ (1998a,b) report fractional
uncertainties in the
convergence $\kappa$ of $\le$ 20--25\%
in the strong lensing regions of interest in the
clusters Abell 370 (using 6 multiple-image systems)
and Abell 2218 (using 3 multiple-imaged systems and a
number of singly imaged, distorted arclets).
They demonstrate
clearly that the mass distribution becomes
more tightly constrained as the number of nearby
lensed images increases.  The dramatic increase in
multiply imaged sources that ACS and later instruments
can uncover should certainly push the uncertainty
in reconstructed cluster mass distributions
well below 10\% in the inner cluster regions.  
Furthermore, strong-lensing features constrain the potential
directly, and
the potential is an integral over the mass
distribution.
Hence we expect the uncertainty in
the cluster potential (the more relevant
uncertainty for $H_0$ determination) to be
less than the uncertainty in the reconstructed
mass distribution.
SN image positions
(and perhaps flux ratios, if microlensing
and differential extinction effects are negligible)
also provide local constraints on the potential,
further improving
the accuracy of a time-delay $H_0$ value.
Taken together these considerations
suggest that statistical $H_0$ errors on the
order of a few percent should be attainable
if successful observations can be made.
And as noted in the introduction,
the lens model will not be subject to
systematic errors associated with a global
mass sheet degeneracy as long as multiple-image
constraints are available for sources
at different redshifts.

To estimate the error that could arise due to
time-delay contributions from
unmodeled galaxy-scale substructure, we
note that time delays scale as the
square of the characteristic deflection of the lens,
which in turn scales as the square of the lens
velocity dispersion, as in (\ref{theta_e})
and (\ref{SIS_diff}).  With a time
delay of $0.1 \tau_{\mathrm{cluster}}$, a
cluster velocity dispersion of
$1000 \, \mathrm{km} / \mathrm{s}$,
and a perturbing galaxy velocity dispersion
of $200 \, \mathrm{km} / \mathrm{s}$,
the induced fractional error in $H_0$
would be
\begin{equation}
{{\delta H_0} \over {H_0}} =
{{\tau_{\mathrm{galaxy}}} \over
{0.1 \tau_{\mathrm{cluster}}}} = 
{{(200)^4} \over {0.1 \times (1000)^4}}
= 1.6 \% ~~.
\end{equation}
Thus substructure is unlikely to be
a source of large error in the
determination of $H_0$ from an observed
time delay.  An appreciable
substructure perturbation to the
overall cluster potential would also
affect lensed SN image positions
(which can themselves constrain the potential),
so it could not go entirely unmodeled by
a reconstruction that makes use
of all lensing constraints.

\section{Predicted Detection Rates}
\label{detect_sec}

We wish to quantify detection rates
for observable SN image pairings
with acceptably short time delays.
Since type II (core-collapse) SNe
will be detected with much higher frequency
than type Ia's \citep[for example]{sul00},
we focus exclusively on the former.
Assuming a Salpeter IMF with lower and upper mass
cutoffs of $0.1 \, M_{\odot}$ and $125 \, M_{\odot}$
respectively, and that all stars with mass
above $8 \, M_{\odot}$ result in core-collapse
SNe, \mbox{\citet{mad98}} derives a conversion factor
of 0.0074 between solar masses of star formation and
eventual number of type II SNe.  The comoving
star formation rate density at a given redshift can in turn
be related to an observable quantity such
as comoving rest-frame ultraviolet
luminosity density, which is seen to rise sharply
out to the redshifts of $z \sim 1$ that we would
hope to probe in our search \citep{mpd98}.

Here we consider two versions of
the cosmic star formation rate density as a function
of redshift.  First, following \citet[``SFH-I'']{sul00},
we use the form of \citet{mp00} corrected to a lower
mass cutoff of $0.1 \, M_{\odot}$ and adjusted from
an Einstein-DeSitter (EdS) universe to our
assumed $\Lambda$CDM cosmology in the manner
described by \citet{hogg01}.
This form incorporates an upward correction to
account for extinction \citep{mad00, mp00}.
We also consider the greatest and least of the
three star formation rate density functions
reported by \citet{lan02} based on their determination
of the star formation rate intensity distribution
function from redshifts $z$ = 0--10, again converted
from EdS to $\Lambda$CDM.  These authors do not
attempt an extinction correction, and we work
directly with their ``unobscured''
star formation rate densities.

We can combine our assumed SN rate densities
with multiple-imaging cross sections
$\sigma(\mu_{\mathrm{min}}, \Delta t_{\mathrm{max}})$
computed numerically for the SIE and NIE
in an integral over source redshifts
to obtain an estimate of the detection rates
that we might expect:
\begin{equation}
\left(
\begin{array}{c}
\mbox{SN image pairings} \\
\mbox{per cluster} \\
\mbox{per obs.\ year}
\end{array}
\right)
= \int_{z_{\mathrm{L}}}^{\infty} {1 \over {1 + z}}
\dot{\rho}_{\mathrm{SN}}(z) \, \sigma(\mu_{\mathrm{min}},
\Delta t_{\mathrm{max}}) \, {{d V_{\mathrm{C}}} \over
{d \Omega \, d z}}(z) \, dz~~.
\end{equation}
$d V_{\mathrm{C}} / d \Omega \, d z$
is the comoving volume per unit solid angle per unit redshift,
the angular cross section depends upon source redshift
and includes a contribution from each independent image
pairing, $\dot{\rho}_{\mathrm{SN}}(z)$ is the rate
density of core-collapse SNe per unit comoving volume per unit proper time,
and the factor of $(1 + z)^{-1}$ converts from proper time at the source
to observer time.  In an actual calculation
this integral must be cut off at some
upper limit beyond which the supernova rate density becomes
utterly unknown.

To obtain a numerical estimate,
we assume a $\sigma_v = 1000 \, \mathrm{km} / \mathrm{s}$
lensing cluster at redshift
$z_{\mathrm{L}} = 0.2$ with an isodensity
axis ratio $q = 0.65$
(as mentioned above, with the exaggerated ellipticity
we hope to approximate the effects of cluster substructure).
We take the maximum tolerable observer-frame time delay
between multiple images to be $3 \, h_{70}^{-1}$ years,
and investigate
a range of differences between limiting
detectable apparent magnitude $m_{\mathrm{lim}}$
and SN II absolute magnitude $M_{\mathrm{SN}}$.
We model the SN II spectrum as that of a
10,000-K blackbody \citep{fil97}, and compute
AB-magnitude
$k$-corrections for observations at 8140$\,$\AA$\,$
\mbox{($I$-band)}.
We consider both an SIE model and an NIE
model with a core radius of
$30h_{70}^{-1} \, \mathrm{kpc}$
(subtending about 9.1 arcseconds).
We normalize the NIE model for
$\overline{z}_{\mathrm{S}} = 0.7$
as described in Section~\ref{NIE_sec}, taking
this as a typical giant-arc
redshift from Williams et al.\ (1999).
Figure~\ref{detect} shows the result of
a detection-rate integration out to a redshift of
$z = 8$.  For visual clarity, we do not plot the
NIE rates for the two
\citet{lan02} star-formation histories;
they show the same increase by a factor of $\sim 2$
over the singular case as do the \citet{mp00} NIE rates.
The results shown are relatively insensitive
to the choice of upper integration limit,
since the
star formation rate density of \citet{mp00}
drops steeply at high redshift, and the
$k$-correction imposes an effective cutoff
on any contribution from the
enhanced high-$z$ star formation
reported by \citet{lan02}.

Although we have considered type II SNe as our
sources, a rough idea of the detection rates for type Ia
SNe image pairings can be obtained
from Figure~\ref{detect} by assuming that
type Ia's are $\sim 10$ times
less frequent and $\sim 1$ magnitude brighter than
type II's.  Under these assumptions,
detection rates for type Ia events are
significantly lower than
for type II's, and the $k$-correction
appropriate to the intrinsically redder colors
of type Ia's will push their
rates lower still.

\section{Outlook}
\label{outlook}

\begin{figure}
\plotone{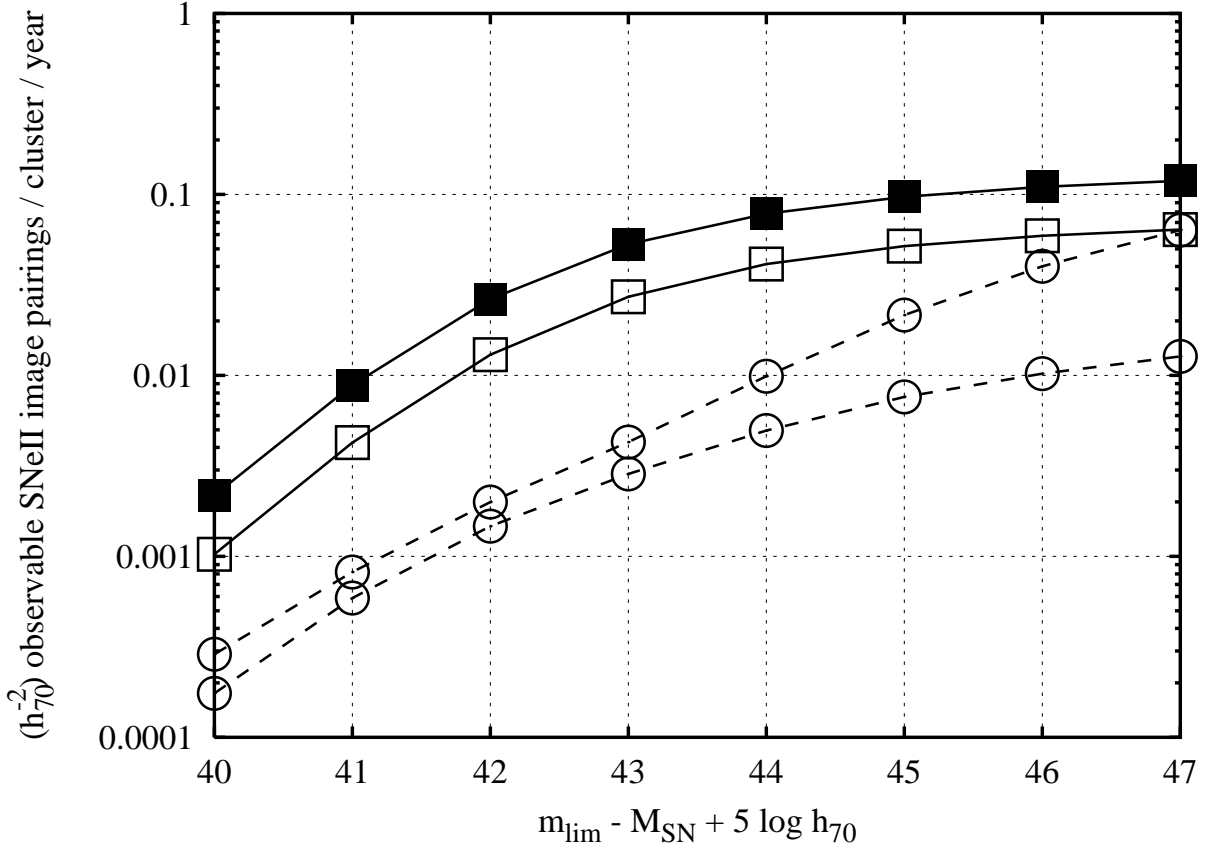}
\caption{Predicted type II SN image pairing detection rates vs.\
difference between limiting observable magnitude and
SN absolute magnitude.
($I$-band, $\sigma_v = 1000 \, \mathrm{km} / \mathrm{s}$,
$q = 0.65$, $z_{\mathrm{L}} = 0.2$,
maximum observer-frame time delay of $3 \, h_{70}^{-1}$ years).
{\em Solid lines/squares:}
star formation rate density of \citet{mp00}.
{\em Dashed lines/circles:} star formation rate density
of \citet{lan02}, upper and lower curves.
Open symbols are for an SIE model; solid symbols
are for an NIE model with a
$30h_{70}^{-1} \, \mathrm{kpc}$ core radius.
\label{detect}}
\end{figure}

\enlargethispage*{1000pt}
The estimated detection rates shown in Figure~\ref{detect}
are far in excess of what one would predict
based on a circular-cluster lens approximation, and
they allow us to gauge the feasibility of a cluster
monitor program targeted to detect multiply-imaged SNe
for use in determining $H_0$.
Repeated imaging would be required every month or so,
as type II-L (II-P)
SNe spend $\sim$30 ($\sim$50) days within one
magnitude of maximum light \citep{db85}.
This time will be stretched by $(1 + z_{\mathrm{SN}})$,
but a precise temporal measurement of peak light
would be crucial to a time-delay
measurement.
Taking $m_{\mathrm{lim}} = 24$
and $M_{\mathrm{SN}} = -18$, the most optimistic
assumptions predict
on the order of 2--3 detectable image pairings
per cluster per century.  A cluster monitor
program operating for two to three years would need
to image $\sim$50--100 cluster fields to the required
depth in order to have a good
chance of detecting a few image pairings.
Such a program would be feasible with
today's 6--8m-class telescopes with wide-field
cameras given several dark nights per month.
The project is better
suited to a telescope such as the
proposed {\sl LSST}, which would operate in a dedicated
survey mode, repeatedly imaging large areas of the sky
(including strongly lensing clusters)
to significant depth.  Although our
predicted image-pair detection rates
are sensitive to assumptions
about cluster structure and cosmic star
formation history,
this method of determining
$H_0$ certainly warrants
serious consideration when such observatories
as {\sl LSST} become operational, particularly
if the current discrepancy between
$H_0$ values from lensing and from other methods
remains unresolved.
\pagebreak

We stress that in order for this method to be
successful, known strong-lensing clusters should be
targeted specifically.  The best constraints
on the cluster potential
would ideally be obtained by follow-up
observations with instruments such as ACS
of clusters for which SN image pairs are
actually observed.
However, the co-added images from a long-term
monitor program would be of significant depth,
and could also be searched for
strongly lensed background features.

\acknowledgments

The authors wish to thank Paul Schechter for valuable
discussion of these topics, Ian Smail for
pointing out the work of \citet{gir92}, and
the anonymous referee for comments that led to a
greatly improved manuscript.

\clearpage

\clearpage

\clearpage

\end{document}